# PubTree: A Hierarchical Search Tool for the MEDLINE Database

*William Rowe[1], Paul D. Dobson[2], Bede Constantinides[3], and Mark Platt[1]*

## Affiliations

1. Department of Chemistry, School of Science, Loughborough University, Loughborough, Leicestershire LE11 3TU, United Kingdom
2. School of Computer Science, The University of Manchester, Oxford Road, Manchester, M13 9PL, UK
3. School of Biological Sciences, The University of Manchester, Oxford Road, Manchester, M13 9PT, UK

## Abstract

Keeping track of the ever-increasing body of scientific literature is an escalating challenge. We present PubTree a hierarchical search tool that efficiently searches the PubMed/MEDLINE dataset based upon a decision tree constructed using >26 million abstracts. The tool is implemented as a webpage, where users are asked a series of eighteen questions to locate pertinent articles. The implementation of this hierarchical search tool highlights issues endemic with document retrieval. However, the construction of this tree indicates that with future developments hierarchical search could become an effective tool (or adjunct) in the mining of biological literature.

## Introduction

As the corpus of scientific literature expands and diversifies our ability to interact with these data in a tractable manner becomes a fundamental challenge. Researchers searching the biological literature commonly engage with the PubMed/MEDLINE dataset using the Entrez search tool[1]. The researcher enters one or more pertinent search terms that then are matched to words found in the associated text of more than twenty six million articles within the database. In this paradigm there are two limiting factors: firstly that the researcher enters a term actually present in the article of interest, and secondly that they have the requisite knowledge of the subject area to choose an appropriate term, use the correct spelling and be aware of other contexts to which that term applies. A study investigating the ability of physicians to answer questions pertinent to one hundred systematic reviews of renal failure found they were able to retrieve fewer than fifty percent of the articles[2]. In addition one in sixteen of the retrieved articles were irrelevant to the search.

An alternative is found in the Medical Subject Headings (MeSH )[3]; a controlled vocabulary used to annotate the PubMed/MEDLINE database. The uses of MeSH are twofold: the indexing of articles, and as a thesaurus to aid search. MeSH are arranged in a hierarchical structure, where levels determine the degree of specificity. In the MeSH structure the hierarchy has been hand constructed through expert knowledge of the relationship between over twenty seven thousand headings and sub-headings to form a twelve level hierarchy. These structures are, however, not optimised in terms of efficiency of hierarchical search. Binary classification trees are graphs where nodes represent questions from which two edges emanate, linking two child nodes characterising the response to each question: yes or no[4]. Modern classification trees such as CART[5] and C4.5[6] are derived from recursive partitioning algorithms, which are commonly used to aggregate large datasets to provide tools for a diverse range of tasks. Whilst these software are often unable to determine the globally most compact representation of the data, they perform consistently well for a range of datasets.

Classification/decision trees are becoming a familiar tool in the repertoire of many biological researchers, where they have been utilised for a wide range of tasks including: characterising signal transduction pathways[7], defining the DNA binding profiles of proteins[8] and predicting patient outcome post-coronary[9]. Our aim is to construct a binary classification tree capable of guiding users through the MEDLINE database to articles of interest via a series of binary questions. Crucial to the performance of a classification tree is the ability to define a set of features that are capable of comprehensively describing the data. We employ a list of over seventy thousand terms with high term frequency–inverse document frequency[10] (terms which separate articles from one another) from PubMed abstracts and titles from recently published highly accessed articles generated by Doğan *et al*[11]. The resultant tree can be interrogated through a bespoke website, forming structured queries to PubMed.

## Methods

### Construction of the Decision Tree

The terms used to build the decision tree were downloaded from the supplementary data of Dogan *et al*. Generic terms (*e.g.* "benefit", "include" and "home") are uninformative in terms of article classification. Such terms were filtered from the list by determining their occurrence in two large non-scientific texts (Moby Dick and War and Peace), in addition to manual filtering (the remaining terms are listed in the supporting information). Initially the terms in the text and the search terms were abbreviated using the Porter stemming algorithm[12] in order to remove common endings such as "ing", "s" *etc*. However, we found searches incompatible with those of PubMed, which does not employ a stemming algorithm. Instead terms were paired with synonyms in PubMed's translation table.

The remaining terms were mapped to the associated text for each article in the MEDLINE database (2016 release downloaded 06/06/2016). Terms were matched: to abstract text, article title, journal title and all associated MeSH terms. This characterisation of the articles forms the raw data upon which the decision tree is built.

$$(x,Y) = (x_1, x_2, x_3..., x_k, Y) \quad [1.1]$$

Where x are the dependent variables (in this case the search terms) and Y is the variable (the articles) we are wishing to classify. The tree is constructed using a simplified recursive partitioning algorithm. Whereby, groups of articles are recursively split into two child groups based on the term (a), which leads to the highest information gain (IG).

$$IG(x,a) = H(x) - H(x|a) \quad [1.2]$$

where $H(x)$ is the entropy before splitting and $H(x|a)$ is the entropy after splitting with term; a, where entropy is defined as:

$$H(x) = -\sum_i P(x_i) \log_2 P(x_i) \quad [1.3]$$

where $P(x_i)$ is the probability mass function of an individual term. To reduce the prevalence of generic terms further down the tree, the entropy was scaled after question four. The entropy of the term in all documents $H(x)^{all}$ was deducted from the entropy of the term in the articles currently assessed $H(x)^{cur}$.

The decision tree described here is far simpler than those used to predict the values of unseen data. As such over-fitting the model to data is not an issue and post-hoc processes such as pruning are unnecessary. Given the size of the dataset processed, the main requisite required in constructing the tree, was to work within a feasible run-time with the smallest memory footprint possible.

## User guided Search

A web interface was developed in JavaScript to allow users to interact with the classification tree. The user is presented with a series of questions, asking whether the current term is found within the sought text, with three answers: "yes", "no" and "maybe". "Maybe" selects the most likely route through the tree based on the remaining papers, but omits the selection from further analysis. The selected search terms are then automatically used to query PubMed either: after eighteen questions, if further questions will fail to refine the search, or earlier at the user's request. Search is enacted through a POST request to "ncbi.nlm.nih.gov", with no expansion of MeSH terms, to more closely match the construction of the decision tree. To lessen the effect of discrepancies between the PubMed search and the mapping applied to construct the tree, UIDs were also included in the searches when end nodes comprised fewer than ten articles. In addition, an option is included to refine the search with bespoke search terms through user entry. Comparisons with standard PubMed searches were also performed with no expansion of MeSH terms, to mirror the performance of PubTree. Searches were performed between 17/10/2016 and 17/11/2016. The tool is accessible from https://bede.github.io/pubtree/.

# Discussion

The recursive partitioning algorithm successfully constructed a hierarchical search algorithm based upon more than twenty six million research articles and thirty five thousand search terms. The terms that arise at the first five levels of nodes in the model are displayed as a tree in Figure 1. "Metazoan" (animals) proves to be the most effective term at splitting the MEDLINE database, as it is found in nearly six million papers. At the next level "biosynthesis" has 5.1 million hits and "adults" has 4.9 million. In total the top four levels of terms are found in 18.8 million documents, the majority of articles in the database. The histogram in Figure 1b displays how the tree subdivided the articles within the database. Ideally, if there were twenty six million articles in the database, the tree would separate them into groups of ca 100 ($26 \times 10^{26} / 2^{18}$) This means that on average each search where the user has answered eighteen questions will (on average) return one hundred articles. The median number of articles retrieved is actually only six, with a wide distribution in the number of articles retrieved. In part this will be due to many articles having little or no metadata associated with them, i.e. no abstract or associated MeSH terms, making it impossible to differentiate them from other articles. For instance, the seminal article describing the structure of DNA by Watson and Crick[13] has no abstract in the database.

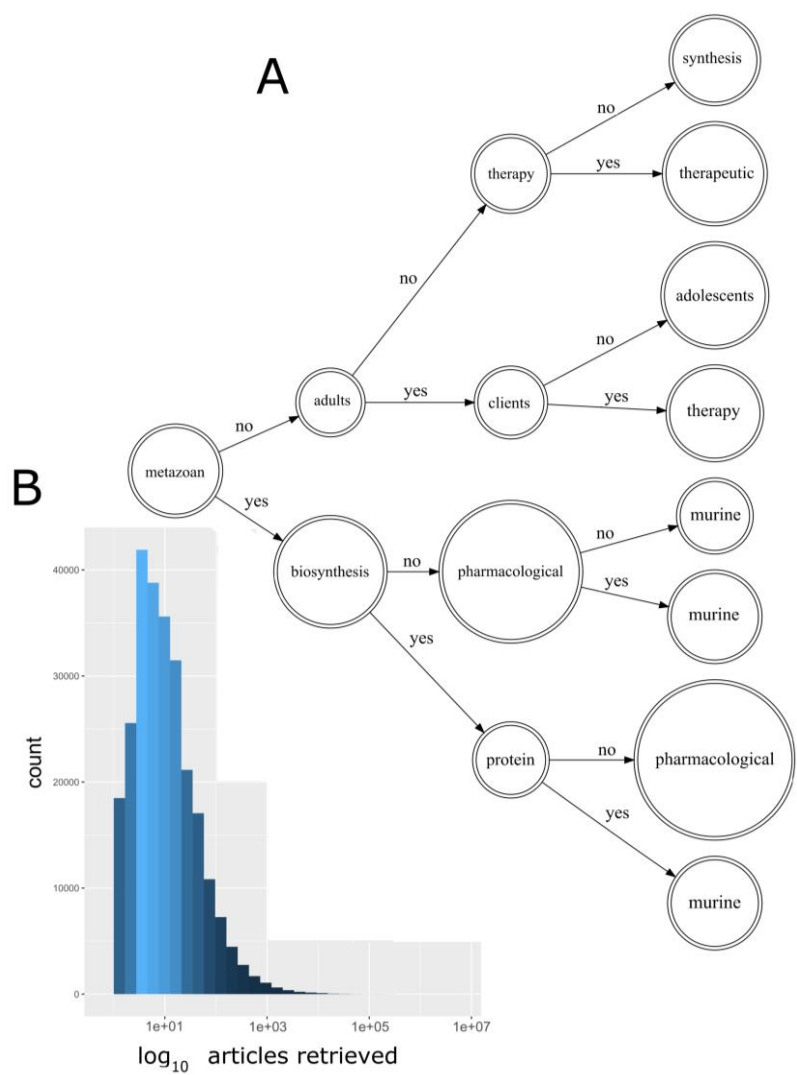

Figure 1. Decision tree structure in PubTree. (A) Graphical representation of the first four levels in the decision tree. (B) Histogram displaying the distribution of the number of articles retrieved by all potential searches with PubTree.

We assessed the performance of PubTree in locating three recent high impact articles from varied fields. Article 1 assesses highly active antiretroviral therapy (HAART) in the treatment of HIV[14], article 2 presents the structure of the eukaryotic ribosome[15], and article 3 is a recent report describing the efficacy of aducanumab in the treatment of Alzheimer's disease[16]. Figure 2 displays the PubTree searches that locate these articles, which also retrieve (respectively) 148, 28 and 189 related articles. The searches begin (as expected) with very generic terms, before focusing on terms specific to the paper. For example, the search for article 1 converges on terms such as "HAART" and "CD4", while the search for article 3 leads to the terms "amyloid" and "plaques". For many purposes retrieving such large numbers of articles will prove impractical. This can be improved by the user supplementing the search with bespoke terms, or in future updates through the expansion of the tree with more questions.

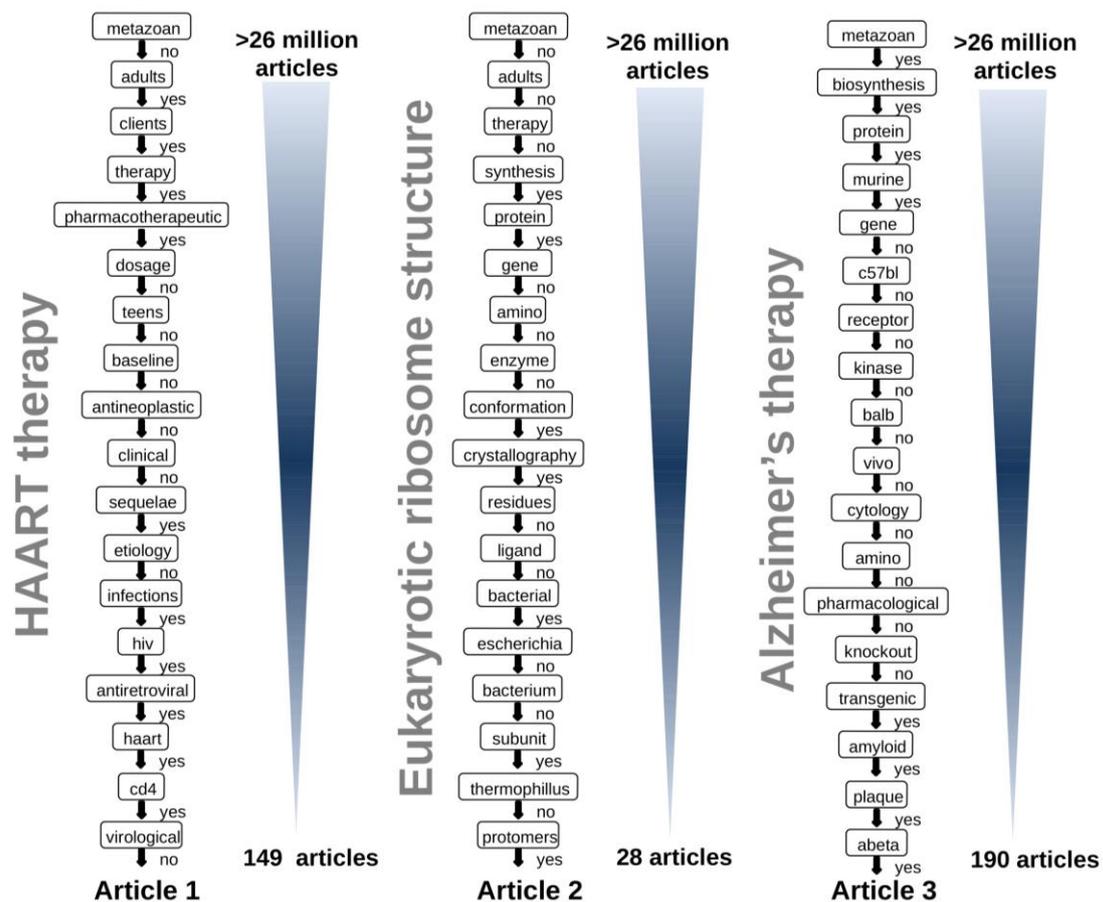

Figure 2. Representation of three searches using PubTree, locating three specified articles. Flow chart representing the questions and responses used to locate three articles: article 1, a description of the efficacy of HAART therapy in the treatment of HIV, article 2, describing the structure of the eukaryotic ribosome, and article 3, presenting the results of an anti-Alzheimer's therapy.

Many limitations exist with the technique. For instance, terms may often be generic, making it difficult for the user to "guess" whether they occur within their target text. In addition terms such as "protein" are ambiguous, as specific proteins may occur in an article but the term "protein" itself may be absent. This might be solved with a more extensive list of synonyms. For example, positive selection of the term "pharmacological" excludes article 3, which may seem counterintuitive for an article describing the positive effects of a new drug therapy. However, this term and surprisingly no associated terms are found in this article. A potential advantage of hierarchical search is that the user can be directed through the search process using a strict controlled vocabulary associated with articles in the database. Therefore, discrepancies between similar terms can be avoided. This and more general use of synonyms will be the subject of further investigation. Building multiple classification trees based on a limited range of publication dates and fields may also improve the efficacy of the approach.

## Conclusions

The tool described in this paper describes a method to efficiently structure predefined terms to aid literature search. We do not envisage the method competing with the currently efficient protocols for interrogating the biological literature; however, the method provides an interesting adjunct to those methods. It may prove effective as an aid to those unfamiliar to a particular field, such as students aiming to survey the literature, or in aiding a researcher locate a difficult to find article. Anecdotally, we have found the tool provides a useful means to identify papers for one of the author's (MP) journal club. In the future we aim to evolve the tool to aid usability by improving synonymous text search, and provide a method for users to give feedback to the partitioning algorithm to improve search performance. Multiple trees can also be constructed to filter by date and broader categories. The tree can be continually updated to include more recently defined search terms with the latest NLM updates, to provide an up to date resource.

## Acknowledgements

The authors gratefully acknowledge helpful discussions with Daniel Jameson and Simon Hubbard.